\documentclass[sigconf,nonacm]{acmart}

\AtBeginDocument{%
  }

\usepackage{graphicx}
\usepackage{natbib}
\usepackage{amsmath,amsfonts}
\usepackage{xcolor}
\usepackage{tabularx}
\usepackage{booktabs}
\usepackage{multirow}
\usepackage{longtable}
\usepackage{acronym}
\usepackage{listings}
\usepackage{stfloats}
\usepackage{algorithm}
\usepackage{array}
\usepackage{xcolor}

\begin{document}

\title{Model-Driven Digital Twin Framework for Quantum Networks}


\author{Amal Elsokary}
\email{A.Elsokary@lboro.ac.uk}
\affiliation{%
  \institution{Loughborough University}
  \city{Loughborough}
  \country{UK}
}

\author{Hayato Ishida}
\email{H.Ishida@lboro.ac.uk}
\affiliation{%
  \institution{Loughborough University}
  \city{Loughborough}
  \country{UK}
}

\author{Ran Wei}
\email{r.wei5@lancaster.ac.uk}
\affiliation{%
  \institution{Lancaster University}
  \city{Lancaster University}
  \country{UK}
}

\author{Michael J. de C. Henshaw}
\email{M.J.d.Henshaw@lboro.ac.uk}
\affiliation{%
  \institution{Loughborough University}
  \city{Loughborough}
  \country{UK}
}

\author{Siyuan Ji}
\email{S.Ji@lboro.ac.uk}
\affiliation{%
  \institution{Loughborough University}
  \city{Loughborough}
  \country{UK}
}

\begin{abstract}
Quantum networks are advancing towards larger and more operational infrastructures, yet their evaluation remains fragmented across heterogeneous physical platforms, simulators, protocols, and architectural abstractions. Current digital-twin studies for quantum networks mainly realise isolated capabilities or application-specific solutions rather than reusable system-level twins. This paper argues that Model-Driven Engineering (MDE) can provide a systematic basis for integrating and evolving these heterogeneous artefacts. It derives requirements for design-time evaluation and runtime synchronisation, and proposes a progression of architectures from code-driven and domain-model-driven solutions to point-to-point and hub-and-spoke integration. A conceptual implementation case study illustrates this using SysML v2, QKD kit, an EMF-based controller, and SeQUeNCe. The work provides a foundation for adaptable and interoperable digital twins for quantum networks.
\end{abstract}



\maketitle

\section{Introduction}
Quantum networks (QNs) are progressing from experimental demonstrations towards increasingly deployed and scalable infrastructures, enabling applications in secure communication, distributed quantum sensing, and distributed quantum computing \cite{QInternet}. This progression is accompanied by continuing advances in quantum photonic devices, communication and entanglement-distribution protocols, network architectures, and control mechanisms. However, relying solely on physical experimentation and deployment to evaluate new technologies and network configurations can be costly and time-consuming and is often constrained by equipment availability, configuration complexity, and experimental risks \cite{martin2024service}.

Moreover, the heterogeneous and interconnected nature of QNs makes it difficult to understand their overall behaviour, predict their performance under emerging conditions, and evaluate the integration of new technologies into existing network environments \cite{lopez2024unleashing}. Their practical adoption therefore requires not only an understanding of individual physical technologies, but also a system-level representation of the interactions and dependencies among components, services, protocols, control functions, and architectural layers \cite{illiano2022quantum}.

These challenges create a need for a structured framework capable of supporting anomaly detection, resource optimisation, technology integration, and the assessment of alternative deployment decisions \cite{lopez2025monitoring}. A digital twin (DT) perspective can address this need by virtually representing network components, services, interfaces, and physical characteristics to explore configurations, analyse behaviour, predict performance, and assess technologies before and during physical deployment \cite{yang2025demonstration}. 


Research on DT for QNs is emerging, with recent studies addressing remote access to experimental data, attack and imperfection analysis, performance optimisation, and monitoring and control of terrestrial and satellite quantum networks \cite{kutschera2021data,martin2024service,ahmadian2024darius,diaz2025digital,yang2025demonstration,mehic2025virtual,chiti2025towards}. These studies commonly employ quantum-network simulators and emulators as executable DT artefacts, including SimulaQron, NetSquid, SeQUeNCe, QKDNetSim, and Qiskit. Collectively, they demonstrate the potential of DT-enabled services, but the field remains at an early stage: most contributions realise selected capabilities or application-specific environments rather than an integrated, system-level DT spanning heterogeneous network representations, simulators, services, and physical assets.

This limitation is reinforced by the absence of a widely adopted QN reference architecture and by the diversity of available simulation tools, which differ in their supported protocols, abstraction levels, fidelity, assumptions, and performance indicators \cite{van2022quantum,lopez2025architectural,elsokarydigital,bel2025simulators}. Consequently, existing models and results are difficult to combine, compare, or reuse across tools and application contexts. The challenge is therefore not only to introduce additional DT functions, but also to establish a systematic engineering basis for composing heterogeneous QN artefacts and progressively realising complete DT capabilities.

Developing such a DT introduces further complexity because models, tools, data sources, and interfaces must remain consistent and traceable across physical and digital environments \cite{erkoyuncu2018digital}. These artefacts operate with different data formats, abstraction levels, timing constraints, and synchronisation rates. Model-Driven Engineering (MDE) provides an opportunity to address this complexity by treating models as explicit and processable artefacts and using metamodels, transformations, model management, and automation to connect system-level representations with domain-specific simulations, monitoring services, and physical-system interfaces \cite{michael2025model}.

Accordingly, this paper proposes an MDE-based framework for engineering QN digital twins at progressively increasing capability levels. It introduces alternative architectures for organising and integrating QN domain models, simulators, services, and physical-system interfaces, and derives the requirements needed to support both design-time evaluation and runtime synchronisation. A feasibility example further illustrates how a SysML v2 system model, an EMF-based simulator controller, and a QN simulator may be connected to evaluate a selected network configuration and its requirements.

\section{Background}

Quantum technology has progressed through two major revolutions \cite{dowling2003quantum,deutsch2020harnessing}. The first quantum revolution, which began in the early twentieth century, established the foundational principles of quantum mechanics, including quantisation, wave–particle duality, and the uncertainty principle. The second quantum revolution emerged from advances in quantum information science, enabling superposition and entanglement to be created, controlled, and exploited as technological resources within individual quantum systems. These developments underpin the fields of quantum communication, quantum computing, quantum simulation, and quantum sensing and metrology \cite{acin2018quantum}.


\subsection{Quantum Networks (QN)}

The development of quantum networks can be traced through the broader evolution of quantum communication, which has become one of the most experimentally mature areas of quantum technology. Early progress was driven largely by secure communication, particularly quantum key distribution (QKD), and this supported the transition from small-scale experiments to operational quantum-network deployments, including the DARPA Quantum Network \cite{etde_20268838}, SECOQC \cite{peev2009secoqc}, and the Tokyo QKD Network \cite{sasaki2011field}. Building on these achievements, quantum networking has expanded across fibre, free-space, and satellite infrastructures, enabling communication over increasingly large distances \cite{liao2017satellite, chen2021integrated}.

Beyond QKD, quantum network encompasses protocols that provide additional security services. Quantum digital signatures (QDS), for example, use quantum states to support message authentication, integrity, and non-repudiation \cite{amiri2015unconditionally}. The broader vision of quantum networking also extends beyond secure communication to the distribution and processing of quantum information through functions such as entanglement distribution, quantum teleportation, blind quantum computation, distributed quantum processing, and high-precision clock synchronisation \cite{kumar2025quantum}, serving as the backbone for scalable quantum computing architectures and the quantum internet. 

As these technologies progress towards practical deployment, research and industrial efforts increasingly focus on integrating the hardware, communication protocols, network architectures, control mechanisms, and operational conditions required for reliable, scalable, and interoperable quantum networks \cite{wehner2018quantum}. However, this progress is hindered by the absence of a unified and standardised quantum-network architecture. Existing standards and architectural proposals adopt different layers, planes, functional components, and software-defined abstractions, making it more difficult to align technologies, coordinate functions, and integrate heterogeneous network elements \cite{stanley2022recent,illiano2022quantum}.

\begin{figure}[ht]
    \centering
    \includegraphics[width=0.6\linewidth]{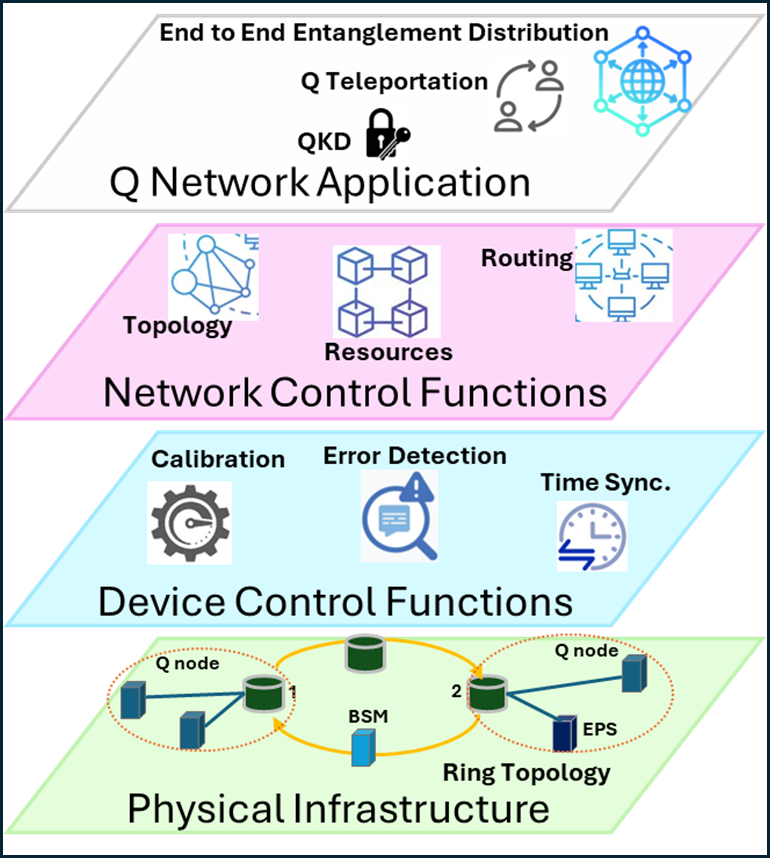}
    \caption{Reproduced architecture of quantum network layers and their functions, developed by IEQNET, led by Fermilab \cite{chung2021ieqnet, chung2022design}}
    \Description{a representative layered quantum-network architecture comprising physical infrastructure, device-control functions, network-control functions, and quantum-network applications. It highlights how application-level capabilities, such as QKD, quantum teleportation, and end-to-end entanglement distribution, depend on coordinated device and network functions across interconnected layers.}
    \label{fig:arch}
\end{figure}

Figure~\ref{fig:arch} presents one layered perspective of a quantum network, comprising physical infrastructure, device-control functions, network-control functions, and quantum-network applications. It illustrates how application-level capabilities, including QKD, quantum teleportation, and end-to-end entanglement distribution, depend on the coordinated operation of devices, protocols, and control functions across interconnected layers.

Although the layers organisation and terminology may differ across architectural proposals, the figure demonstrates that the operation and performance of each layer depend on functions, resources, and information exchanged across other layers. This architectural diversity, together with the interdependence of network technologies, complicates the integration of heterogeneous components, the alignment of functions across layers, and the evaluation of system-level performance. Digital twins (DTs) offer a promising approach to addressing these challenges by providing connected digital representations through which layered architectures can be modelled, evaluated, and validated. They can also support the exploration of alternative configurations, the analysis of system behaviour, and operational decision-making.

\subsection{Digital Twin (DT)}

Digital twins (DTs) are being adopted across cyber-physical system (CPS) domains to support capabilities such as anomaly detection, virtual experimentation, performance evaluation, and adaptive planning \cite{wei2024towards, lehner2025model}. However, there is no single agreed definition or architecture for a DT. Its requirements, functions, components, and system boundaries are typically defined according to the objectives and characteristics of the application domain \cite{dalibor2022cross}. As a result, numerous DT architectures have been proposed, with limited agreement on their common components, functional elements, and interfaces \cite{nwogu2022towards}.

\begin{figure}[ht]
    \centering
    \includegraphics[width=1\linewidth]{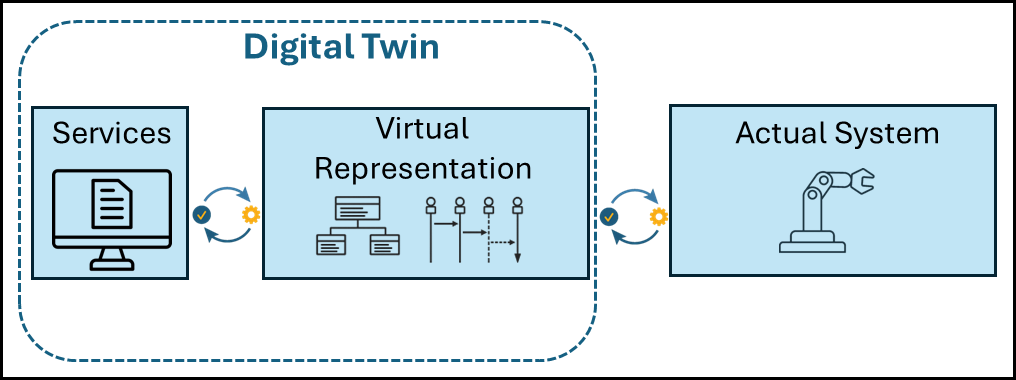}
    \caption{DT conceptualisation to three main components: actual system, virtual representation (mirror), and service for synchronisation or user interface \cite{michael2025model}}
    \Description{A diagram showing the three main components of a digital twin: actual system, virtual representation, and service for synchronisation or user interface.}
    \label{fig:DT}
\end{figure}

Rather than proposing a new definition, this study adopts the architectural conceptualisation of Michael et al. \cite{michael2025model}, in which a DT comprises three main components, as illustrated in Figure~\ref{fig:DT}. The first is the actual system operating in the real environment and exchanging the data required for synchronisation. The second is the virtual representation, comprising models that describe the configuration and current state of the actual system. The third is a set of services supporting synchronisation at a defined rate, analysis derived from models, and the visualisation and reporting of information to DT users.

Accordingly, a DT is considered a software-intensive system that connects an actual system with its virtual representation and provides services for monitoring, analysis, simulation, prediction, and decision support \cite{michael2025model}. Depending on its purpose, it may also provide feedback that influences the state or behaviour of the actual system. DT-related models can be distinguished as models \textit{in} the DT and models \textit{of} the DT \cite{michael2025model, hellwig2025digital}. Models in the DT represent selected aspects of the actual system, such as its structure, behaviour, parameters, and operational state, and may include physics-based, simulation, or data-driven models. Models of the DT represent the DT itself as a software-intensive system, including its architecture, services, interfaces, and synchronisation mechanisms. This complexity gives rise to a fundamental engineering challenge: how DTs can be designed, developed, and evolved efficiently, effectively, and systematically \cite{michael2025model, zhang2025digital}.

\subsection{MDE for Quantum Network-DT}

The combined complexity of representing quantum networks and engineering their digital twins creates a need to coordinate heterogeneous models, tools, interfaces, and configurations. Model-Driven Engineering (MDE) provides a suitable foundation by treating models as first-class artefacts and using abstraction and automation to support the development, integration, maintenance, and evolution of software-intensive systems \cite{brambilla2017model,da2015model,schmidt2006model}.

In a quantum-network DT, modelling languages and metamodels can formalise system concepts, configurations, and interface contracts, while transformations, code generation, and model interpretation can connect system-level representations with domain-specific simulators, runtime services, monitoring mechanisms, and physical-system interfaces. Multi-view modelling, traceability, and consistency management can preserve relationships across structural, behavioural, physical, and operational representations, while variability mechanisms support the systematic selection and comparison of alternative protocols, technologies, simulators, and network configurations \cite{cicchetti2019multi,schmidt2006model,brambilla2017model}. Collectively, these techniques make integration decisions explicit and enable changes to be propagated across otherwise heterogeneous artefacts.

These capabilities can be applied both to models \textit{in} the DT, which represent the quantum network, and to models \textit{of} the DT, which define its architecture, services, interfaces, and synchronisation mechanisms. MDE therefore provides a structured basis for integrating, validating, and evolving both the network representation and the supporting DT infrastructure throughout their lifecycle \cite{lehner2025model,burgueno2025automation,michael2025model}. Although it does not remove all DT engineering challenges, it offers established mechanisms for addressing them systematically.

\begin{figure}[ht]
\centering
\includegraphics[width=\linewidth,
trim={2cm 2cm 2cm 2cm},
    clip
    ]{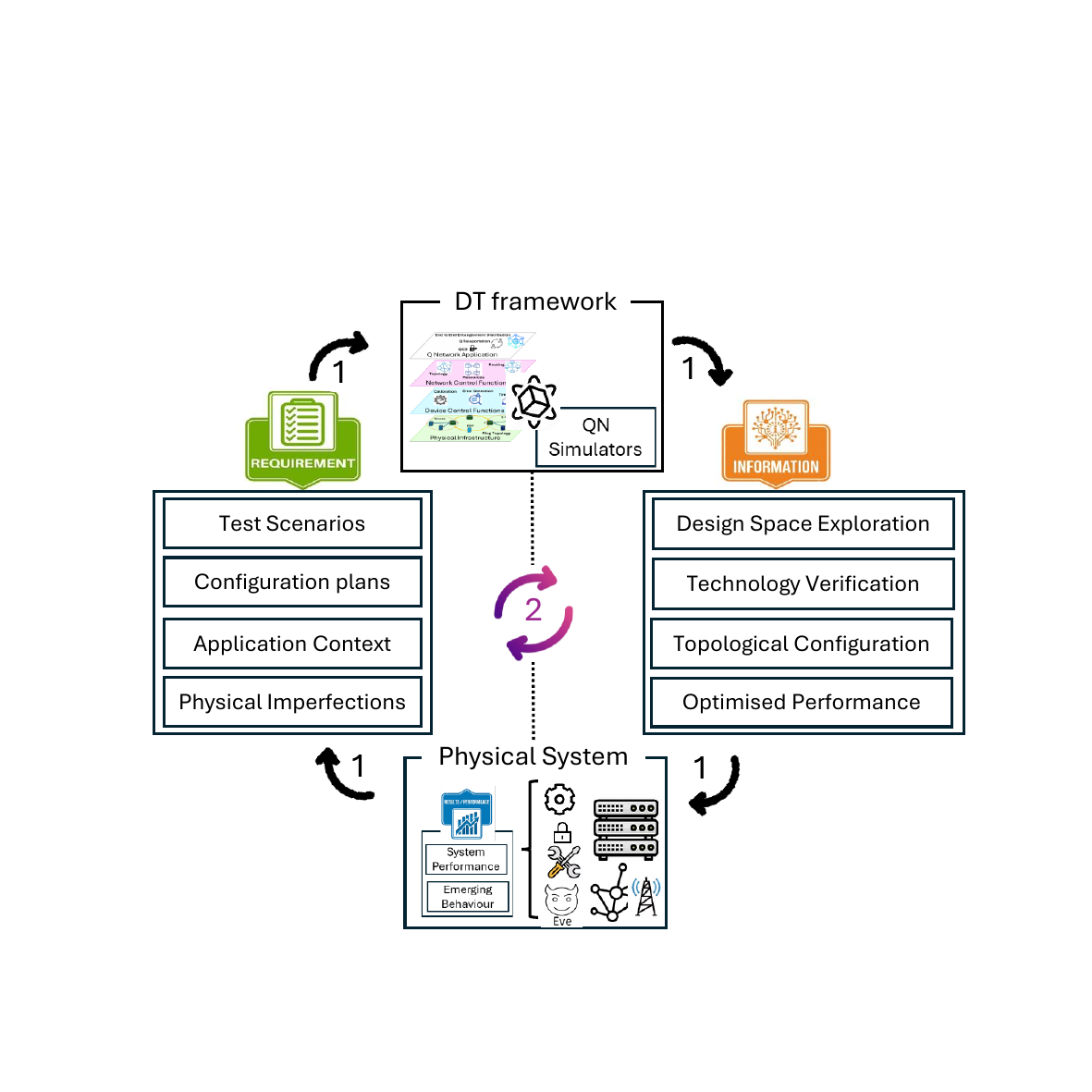}
\caption{DT framework requirements across the design-time and runtime loops.}
\Description{Engineering requirements for a quantum-network digital twin framework, represented through the design-time engineering loop and the runtime synchronisation and co-evolution loop.}
\label{fig:Req}
\end{figure}

The framework requirements are organised around the two complementary loops shown in Figure~\ref{fig:Req}. Loop~1 represents the design-time engineering process, in which requirements, test scenarios, configuration plans, application contexts, and physical imperfections are used to configure suitable QN models and simulators. The resulting information supports design-space exploration, technology verification, topological configuration, and performance optimisation, and may subsequently inform the implementation or reconfiguration of the physical system.

Loop~2 represents runtime synchronisation and co-evolution between the DT framework and the operational physical system. Measurements, system states, performance indicators, and emergent behaviours update the digital representations at an appropriate synchronisation rate, enabling comparison between predicted and observed behaviour, deviation detection, and assessment against operational requirements.

Together, the two loops define the requirements for a QN-DT framework that supports both the evaluation of alternative configurations before deployment and the continued monitoring and evolution of the network during operation. The following sections present the MDE-based architectures proposed to realise these requirements.

\section{Proposed Approaches}
\begin{figure*}[ht]
    \centering
    \includegraphics[width=\linewidth]{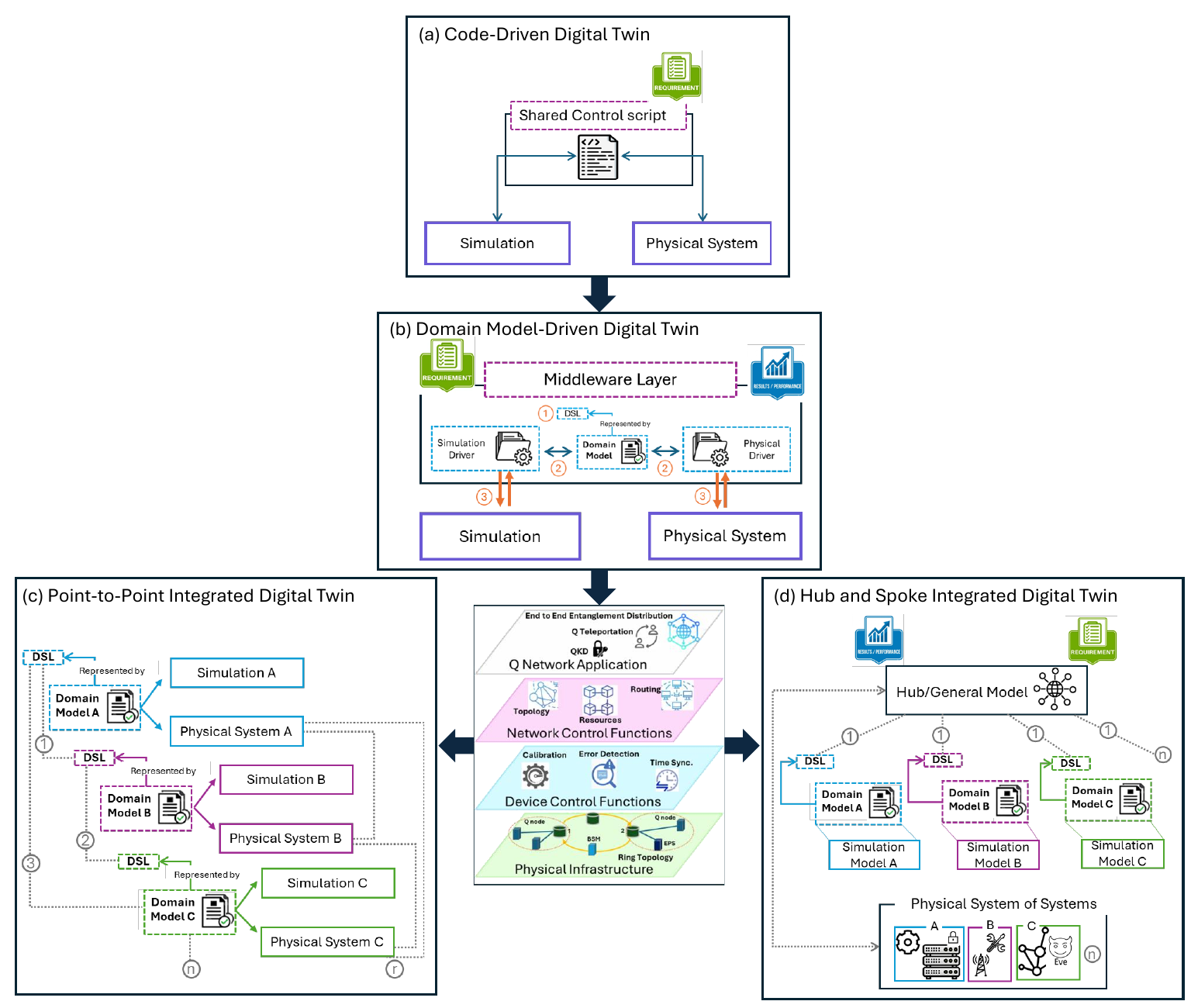}
    \caption{Progressive architectures for QN digital twins: (a) a script-driven architecture in which a shared control script directly coordinates the simulated and physical systems; (b) a domain-model-driven architecture in which a domain-specific model and embodiment-specific drivers govern their configuration and interaction; (c) a point-to-point integrated architecture in which multiple DSLs interact through direct relationships to represent cross-domain interdependencies; and (d) a hub-and-spoke integrated architecture in which a general system model coordinates multiple domain models and the corresponding physical QN system of systems.}
    \Description{The four proposed architectures are illustrated as a progression from implementation-specific coupling, to domain-model-based integration, and finally to system-level coordination of multiple QN representations.}
    \label{fig:proposed}
\end{figure*}

Figure~\ref{fig:proposed} presents four progressive architectures for engineering digital twins of quantum networks using MDE techniques. The architectures differ in both the artefact used to establish the relationship between the simulated and physical systems and the scope of the QN being represented. In Architecture~(a), configuration and synchronisation knowledge is embedded in a shared control script. Architecture~(b) makes this knowledge explicit through a domain-specific model and its associated simulation and physical drivers. Architectures~(c and d) extend this model-driven approach to a QN system of systems containing multiple interacting domains, simulators, and physical subsystems. Two alternatives are considered at this level: direct point-to-point integration among the participating domain models and integration through a common system model. The progression therefore moves from implementation-specific code, through single-domain model-driven integration, to multi-domain system-level integration.

\subsection{ Code-Driven Digital Twin}

In this architecture, shown in Figure~\ref{fig:proposed}(a), a bespoke shared control script directly coordinates a predefined simulated--physical-system configuration. The script contains the parameter mappings, execution logic, data conversions, and feedback rules required to configure both environments, initiate their execution, collect their outputs, and compare their behaviour. Its implementation is feasible through software interfaces and embedded control hardware, such as FPGA-based controllers, which can translate script commands into physical actions and acquire measurements from the experimental system. This provides a practical means of applying corresponding configurations to the simulator and hardware and assessing their performance.

However, the architecture does not provide an automated data or feedback pipeline between the two environments. The user must observe and interpret the results, determine the next action, and manually initiate a further execution of the script. It therefore supports a human-in-the-loop sequence of experiments rather than continuous synchronisation or autonomous closed-loop control.

Its limitation is that the integration knowledge remains embedded in implementation-specific code and is closely coupled to the available parameters, sensing channels, hardware interfaces, data formats, and commands. Introducing an additional property for evaluation may require not only revision and revalidation of the script, but also extensions to the physical measurement and control interfaces. One example is provided by Ahmadian et al. \cite{ahmadian2024darius}, extending an experiment from evaluating channel attenuation to monitoring and compensating for state-of-polarisation variations requires additional measurements, optical-component controls, and interpretation logic to relate polarisation changes to QBER and key-exchange performance. Consequently, adopting the architecture for a different QN configuration or evaluation concern may require coordinated modification of the script and both environment-specific interfaces.


\subsection{Domain Model-Driven Digital Twin}

The second architecture, illustrated in Figure~\ref{fig:proposed}(b), introduces a domain-specific modelling layer for quantum network systems. The domain model forms part of a middleware layer that coordinates the simulated and physical environments through simulation and physical drivers. These drivers translate domain concepts into environment-specific configurations, commands, and data. MDE techniques enable this coordination: model interpretation allows behavioural or orchestration models to be executed directly, while model-to-model and model-to-text transformations derive simulator configurations, interface adapters, hardware commands, and other executable artefacts required to operate both environments. Validation rules may additionally be used to check constraints and parameter mappings before execution.

The progression from the code-driven to the domain-model-driven architecture reflects a change in how system and integration knowledge is represented, processed, reused, and maintained. The code-driven architecture is primarily suited to a predefined experimental configuration. A change in the network topology, protocol, component parameters may require the script to be modified manually and subsequently realigned with both environments. The domain model-driven architecture introduces a higher level of abstraction above this implementation-specific coordination. The domain model does not merely provide a descriptive representation of the QN. Behavioural and orchestration models are inetnded to be processable and executable, enabling information and data exchange to be derived systematically for both the simulated and physical environments. This enables the simulated and physical representations to evolve together while preserving their correspondence with the authoritative model.

\subsection{System-of-Systems Model-Driven Digital Twin}

Architecture~(b) establishes a model-driven relationship between a domain model, a quantum-network simulator, and the corresponding physical system. This is appropriate where the DT is intended to evaluate a particular QN concern or a selected simulator--physical-system pairing. However, the behaviour and performance of a quantum network as a whole cannot always be determined from one domain in isolation. A QN contains interacting physical and software-intensive subsystems associated with concerns such as quantum devices and channels, protocol operation, security, routing, scheduling, resource management, and classical control. The behaviour represented in one domain may therefore influence the assumptions, parameters, and outputs of other domains.

Architecture~(c and d) extends the domain-model-driven approach to this system-of-systems scope. Multiple specialised domain models are used to represent the relevant QN concerns, while each remains associated with the quantum-network simulator and physical subsystem appropriate to that domain. MDE is required not only to derive executable artefacts for each simulated--physical pairing, but also to represent, transform, and maintain the dependencies among the participating domain models. These dependencies are necessary for evaluating the integrated behaviour of the network, because effects introduced within one subsystem may propagate through other subsystems and influence system-level measures such as QBER, secret-key rate, latency, and service delivery.

Two model-driven integration alternatives are considered. Figure~\ref{fig:proposed} (c) illustrates the domain models as independent authoritative representations and establishes direct point-to-point mappings, transformations, and consistency mechanisms between those that interact. Figure~\ref{fig:proposed} (d) introduces a common QN system model through which the participating domain models and their dependencies are coordinated. Both alternatives represent the same physical system of systems and preserve the specialised simulator associated with each domain; however, they differ in how cross-domain integration knowledge is organised and how the required engineering effort grows as additional domains are incorporated.

To formalise this comparison, let the physical QN system of systems be represented as

\begin{equation}
G_{P}=(V_{P},E_{P}),
\end{equation}

where $V_{P}=\{P_{1},P_{2},\ldots,P_{n}\}$ denotes the participating physical subsystems and $E_{P}$ denotes their physical and operational interdependencies. Each subsystem or engineering concern is represented by a domain model $M_i$, constructed using a DSL. Each domain model is associated with its relevant simulator $S_i$ and physical subsystem $P_i$:

\begin{equation}
M_i \models DSL_i,
\qquad
M_i \leftrightarrow S_i,
\qquad
M_i \leftrightarrow P_i,
\qquad
i=1,\ldots,n.
\end{equation}

The following two alternatives examine how the dependencies among these domain models can be engineered to support evaluation of the integrated QN behaviour.

\subsubsection{Point-to-Point Integrated Digital Twin:}
\leavevmode\par
\noindent

In the point-to-point approach, each domain model remains an independent representation of its corresponding physical subsystem and simulation environment, as shown in Figure~\ref{fig:proposed}(c). The integrated behaviour of the network is established through direct relationships between the participating DSLs. In the illustraitve example in Figure~\ref{fig:proposed}(c), each physcial system of the three is interfacing every other physical system, resulting in the need of 3 relationships to be established between the corresponding DSLs. Depending on the nature of the relationships, they might be implemented using model-to-model transformations, model interpretation, semantic mappings, and code generation.

Depending on the actual need for DSL relationships to consistently reflect those appearing between the physical systems, for $n$-DSLs, the total number of unique pairwise relationships between is bounded by a worst case number, calculated as

\begin{equation}
r(n) = \sum_{i=1}^{n-1} r_i = \frac{n(n-1)}{2},
\end{equation}

suggesting a qudratic scaling, as new system is introduced into the network. 

In addition, every domain model must remain connected to its corresponding simulator and physical subsystem. Consequently, modifying or extending one domain model, or its governing DSL metamodel, may require several related mappings and interfaces to be revised and revalidated. This approach remains feasible for DTs with a limited scope, involving only a small number of domain models or DSLs that share a common metamodel and have stable dependencies. However, its maintainability and scalability decrease as the DT expands to represent wider quantum networks with multiple interacting domains, and could quickly become infeasible for a real-world system of systems that has multi-layers and multi-systems, like the one shown in Figure~\ref{fig:proposed}.

\subsubsection{Hub-and-Spoke Integrated Digital Twin:}
\leavevmode\par
\noindent

In the hub-and-spoke approach, the specialised domain models do not exchange information through direct pairwise mappings. Instead, each domain model is related independently to a general QN system model, which acts as the hub through which their concepts, parameters, results, and interdependencies are integrated. Each domain model remains associated with the quantum-network simulator appropriate to its concern, while the hub model provides the system-level representation required to interpret their combined effects on the operational physical network.

At the model level, the corresponding domain model $(M_i)$ is transformed, projected, or interpreted within the general QN system model $(M_QN)$. Each domain model also retains a one-to-one association with its corresponding simulator $(S_i)$:

\begin{equation}
\mu_i \rightarrow M_{\mathrm{QN}},
\qquad
M_i \leftrightarrow S_i.
\end{equation}

When domain models are introduced sequentially, each additional model requires only one mapping to the hub. Therefore, for $n$-domain models, the total number of domain-to-hub mappings is

\begin{equation}
N_{\mathrm{Hub}}(n) = n.
\end{equation}

Accordingly, the number of required mappings grows linearly with the number of participating domain models. Each domain model additionally requires one association with its corresponding simulator. Under the simplifying assumption that the hub model maintains one system-level direct relationship with the operational physical QN.

One scenario implementation may integrate two domain models. For example, a QKD protocol model may represent basis selection, sifting, and key generation, while a channel model may represent Eve's interception, attenuation, and noise. Integrating these models through the hub enables their combined effect on QBER and secret-key rate to be assessed.

\begin{table*}[ht]
\centering
\caption{Capabilities and limitations of the proposed architectures against the four DT capabilities}
\label{tab:4Rs}

\renewcommand{\arraystretch}{1.25}
\setlength{\tabcolsep}{3pt}
\small

\begin{tabular}{|p{2.6cm}|p{3.4cm}|p{3.4cm}|p{3.4cm}|p{3.4cm}|}
\hline

\textbf{Architecture} &
\textbf{R1: Representation} &
\textbf{R2: Replication} &
\textbf{R3: Reality} &
\textbf{R4: Relation} \\
\hline

\textbf{Architecture ~a: Code-Driven Digital Twin} &
\textbf{Limited:} Represents one predefined, implementation-specific configuration. &
\textbf{Limited:} Applies corresponding parameters to the simulated and physical systems and compares their outputs. &
\textbf{Limited:} Supports prediction and assessment, but new scenarios require additional coding and manual modification. &
\textbf{Not supported:} Synchronisation, feedback, and updates are not supported. \\
\hline

\textbf{Architecture ~b: Domain Model-Driven Digital Twin} &
\textbf{Supported:} Represents structure, behaviour, requirements, constraints, and relationships within one domain. &
\textbf{Supported:} Derives simulated and physical configurations from the same domain model. &
\textbf{Supported:} Enables reusable scenarios and configurations through interpretation and transformation. &
\textbf{Limited:} Runtime synchronisation requires external drivers and execution infrastructure. \\
\hline

\textbf{Architecture ~c: Point to Point Integrated Digital Twin} &
\textbf{Supported:} Represents multiple QN subsystems and their pairwise dependencies. &
\textbf{Supported:} Replicates subsystems through corresponding simulations and physical mappings. &
\textbf{Limited.} System-level analysis depends on the correct coordination of the participating models and simulations. &
\textbf{Limited.} Growing mappings and consistency rules complicate synchronisation and feedback. \\
\hline

\textbf{Architecture ~d: Hub and Spoke Integrated Digital Twin} &
\textbf{Supported:} Represents the QN as an integrated system of specialised domain models. &
\textbf{Supported:} Combines simulation outputs to replicate physical QN behaviour. &
\textbf{Supported:} Enables system-level analysis, optimisation, and cross-domain assessment. &
\textbf{Supported:} Enables system-level synchronisation and feedback, subject to reconciling timing, abstraction, assumptions, and fidelity. \\
\hline

\end{tabular}
\end{table*}

A wider implementation may include a third model representing device imperfections, such as detector efficiency, dark counts, source intensity, or polarisation drift. A fourth model may represent network-level concerns, including topology, routing, scheduling, resource allocation, or key-management demand. The hub model relates these domains to evaluate how device behaviour, protocol operation, channel conditions, and network-control decisions collectively affect system-level performance.

The dependencies represented within the hub model correspond to the interactions that already exist among the physical subsystems in the operational QN. The hub model does not remove these physical dependencies. Instead, it provides a common representation through which their propagated effects can be traced, analysed, and related to physical measurements and control actions.

The novelty of this proposed MDE-enabled approach lies in providing a reusable system-level structure through which the DT can be adapted to different QN configurations, simulators, and engineering artefacts, rather than remaining limited to isolated and purpose-specific evaluations.

These architectures represent increasing degrees of model governance rather than mutually exclusive definitions of a digital twin. Script-based adapters may remain necessary at the physical and simulator boundaries, while domain-specific models provide specialised executable representations. The proposed contribution is to govern these artefacts through a system-level model that makes their configuration, relationships, transformation paths, and physical counterparts explicit across the progressive DT capabilities, as discussed in table~\ref{tab:4Rs}.

\section{Proof of Concept}

As discussed earlier, an adaptable and reusable framework is needed to address the diversity and complexity of quantum network systems. This section presents a proof of concept for selected elements of the hub-and-spoke architecture introduced in Architecture~(d). The purpose is to demonstrate a feasible implementation path in which the physical hardware configuration and system requirements are represented in the hub model, translated into simulator-specific configuration information, and executed by a selected simulator.

SysML v2 is selected for the hub model because it provides graphical and textual modelling capabilities \cite{omgSysMLv2}, supports the distinction between definitions and usages, and offers a standard API for integration and automation \cite{omgSystemsModelingAPI}. Its graphical notation can represent the physical network, including its structural and behavioural relationships, while its machine-processable textual notation includes supports model querying, transformation, and integration with external tools. Moreover, the distinction between definitions and usages enables reusable QN concepts to be instantiated and specialised for particular protocols, physical configurations, and evaluation scenarios. The standard modelling API further supports interoperability and automation across connected artefacts.

\begin{figure}[ht]
    \centering
    \includegraphics[width=1\linewidth]{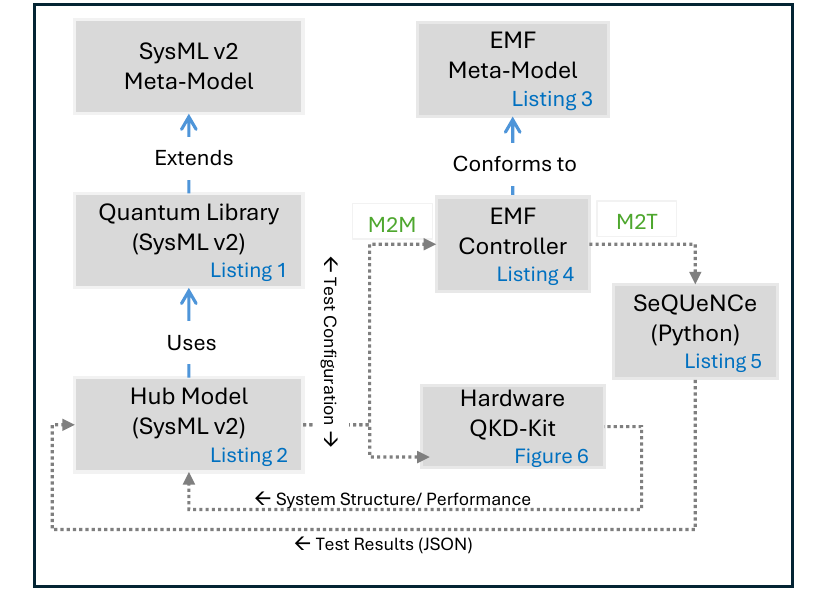}
    \caption{Mapping of the hub-and-spoke QN digital-twin architecture to the BB84 feasibility implementation.}
    \Description{A stacked implementation diagram showing the SysML v2 hub, EMF controller, SeQUeNCe simulator, and physical BB84 QKD kit.}
    \label{fig:Imp}
\end{figure}

The path in Figure~\ref{fig:Imp} illustrates a hub model instantiated according to the physical structure of the system. The hub model defines the required system structure, evaluation requirements, and test configurations, which can then be tested through both simulation and physical environments. Integration with the simulator is mediated through an EMF-based domain-specific controller.

A model-to-model transformation uses the SysML v2 configuration to create a simulator-control model. This separation keeps the hub independent of a particular simulator and enables alternative simulators to be integrated through different domain-specific controllers and adapters. Finally, the simulated and measured results from both environments are returned to the hub model and evaluated against the defined requirements to check compliance. Direct integration between the hardware and SysML environment can be supported through a hardware-interface mechanism, such as that demonstrated by SHIA \cite{lewis2026shia}.

Listing~\ref{lst} illustrates the quantum-network library, where concepts and definitions are modelled for further instantiation and specialisation. Application-specific QN packages are defined to represent selected network implementations and evaluation contexts. The \texttt{SimulationExchangePort} and \texttt{HardwareMeasurementPort} definitions prepare the hub model to exchange configurations and results with external artefacts, including simulator controllers and physical hardware. These definitions are therefore not intended to represent one specific setup directly, but to provide reusable modelling concepts from which later QKD configurations can be instantiated and specialised.

\begin{lstlisting}[
caption={SysML v2 representation of the general QN hub model.},
label={lst},
frame=single,
basicstyle=\ttfamily\footnotesize
]
package QN {
    package QNApplicationContext {
        package QKD_Network {
            part def QKDprtocol;
            part def sender;
            part def receiver;
            connection def QuantumChannel;
            connection def ClassicalChannel;
            enum def QKD_Protocol {
                BB84;
                B92;
            }
            item def testConfiguration;
            item def PerformanceResult {
                attribute throughput : Real;
                attribute keyRate : Real;
                attribute latency : Real;
                attribute errorRate : Real;
            }
            requirement def performanceRequirement {
                subject performance : QKD_Protocol;
                attribute maxValue : Real;
            }
            part def PhysicalQKDSystem;
            part def QuantumNetworkSimulator;
            part def SimulatorController;
            port def SimulationExchangePort;
            port def HardwareMeasurementPort;
        }
    }
}
\end{lstlisting}

\begin{figure}[ht]
    \centering
    \includegraphics[width=1\linewidth]{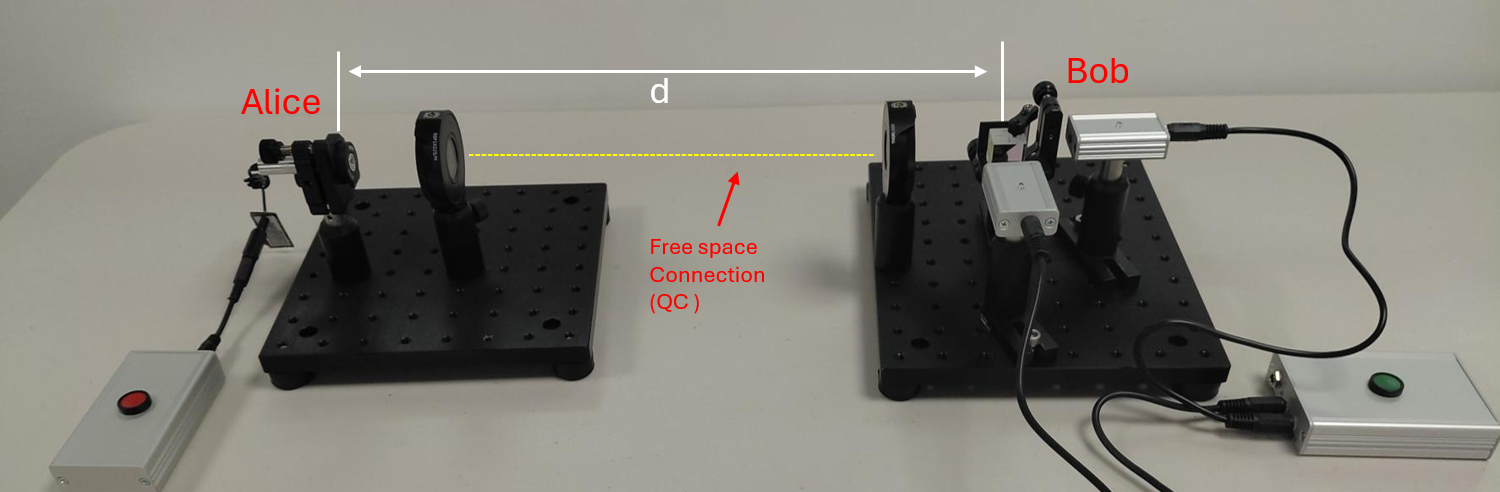}
    \caption{BB84 QKD hardware}
    \Description{Alice and Bob communicate through quantum and classical channels to generate a shared key.}
    \label{fig:HW}
\end{figure}

This feasibility example uses a BB84 QKD kit as a simple setup to demonstrate the implementation feasibility, as shown in Figure~\ref{fig:HW}. Alice and Bob are connected through a free-space quantum channel with distance \textit{d}, and the two nodes communicate to generate a secret key. To produce the final key, a post-processing step is required, in which the error rate is measured as an indicator of performance. The distance \textit{d} is treated as a configurable parameter that can be varied to evaluate its effect on the resulting error rate.

Based on this experiment, a SysML hub model is created to act as a single source of truth in which the actual system and DT configuration are represented. This enables network configurations, requirements, test scenarios, interface contracts, and returned performance results to be captured in a single model, as shown in Listing~\ref{2nd}. The Alice and Bob part usages, their quantum and classical channel usages, and the test-configuration usage are typed by the definitions introduced in Listing~\ref{lst}.

The \texttt{controllerPort} and \texttt{hardwarePort} support the respective exchanges with the selected simulator and physical kit, while the \texttt{errorRate} requirement usage defines the threshold applied to both returned results. The simulated and measured error-rate values are allocated to their corresponding elements and evaluated against the defined error-rate requirement. A configuration is accepted when both constraints are satisfied; otherwise, additional parameters, such as attenuation or detector characteristics, can be incorporated and evaluated through both environments.

\begin{lstlisting}[
caption={SysML v2 representation of the BB84 configuration, exchange ports, and error-rate requirement.},
label={2nd},
frame=single,
basicstyle=\ttfamily\footnotesize
]
package BB84_Network {
    private import QNApplicationContext::QKD_Network::*;
    part BB84_Protocol : QKDprotocol {
        part Alice : sender;
        part Bob : receiver;
        connection QC : QuantumChannel
            connect Alice to Bob {
            attribute distance : Real;
        }
        connection CC : ClassicalChannel
            connect Alice to Bob;
    }
    requirement errorRate: performanceRequirement {
        attribute maxAllowedErrorRate=0.11;
        require constraint {
            simulatedErrorRate <= maxAllowedErrorRate
        }
        require constraint {
            measuredErrorRate <= maxAllowedErrorRate
        }
    }
    item QC_configuration : testConfiguration {
        attribute QC_distance = 1 [m];
    }
    item measuredErrorRate : PerformanceResult;
    item simulatedErrorRate : PerformanceResult;
    part qkdKit : PhysicalQKDSystem;
    part SeQUeNCe : QuantumNetworkSimulator;
    part emfController : SimulatorController;
    port controllerPort : SimulationExchangePort {
        out item QC_configuration : testConfiguration;
        in item simulatedResult : PerformanceResult;
    }
    port hardwarePort : HardwareMeasurementPort {
        out item QCconfiguration : testConfiguration;
        in item measuredResult : PerformanceResult
    }
}
\end{lstlisting}

A model-to-model (M2M) transformation uses the SysML v2 configuration and evaluation scenario to create a simulator-control model. The controller metamodel defines the structure that this model must conform to, including the simulator and experiment to be executed, the input parameters to be supplied, and the outputs to be returned. The resulting controller instance provides an explicit mapping between SysML v2 properties and the corresponding simulator inputs and outputs. Listing~\ref{3rd} presents the controller metamodel using Emfatic.

\begin{lstlisting}[
caption={Emfatic metamodel for the simulator controller.},
label={3rd},
frame=single,
basicstyle=\ttfamily\footnotesize
]
@namespace(
uri="http://qndt/controller",
prefix="qnctrl"
)
package controller;
class SimulatorController {
attr String simulator;
val Experiment experiment;
val Input[*] inputs;
val Output[*] outputs;
}
class Experiment {
attr String protocol;
attr String sender;
attr String receiver;
}
class Input {
attr String sysmlQualifiedName;
attr String simulatorParameter;
attr EDouble value;
attr String unit;
}
class Output {
attr String simulatorMetric;
attr String resultFile;
attr String sysmlQualifiedName;
}
\end{lstlisting}

Listing~\ref{4th} presents an XMI controller instance that invokes a SeQUeNCe simulator \cite{wu2021sequence} and BB84 experiment, and maps the SysML v2 quantum-channel distance to the corresponding simulator parameter. The output binding maps the \texttt{mean\_error\_rate} field in the JSON result to the corresponding simulated error-rate property in the SysML v2 model.

\begin{lstlisting}[
caption={EMF controller instance for the SeQUeNCe BB84 experiment.},
label={4th},
frame=single,
basicstyle=\ttfamily\footnotesize
]

<?xml version="1.0" encoding="ASCII"?>

<qnctrl:SimulatorController
xmi:version="2.0"
xmlns:xmi="http://www.omg.org/XMI"
xmlns:qnctrl="http://qndt/controller"
xmi:id="_NS7_4HbKEfGv-tGzRUrLbQ"
simulator="SeQUeNCe">
<experiment
    xmi:id="_QC9QcHbKEfGv-tGzRUrLbQ"
    protocol="BB84"
    sender="Alice"
    receiver="Bob"/>

<inputs
    xmi:id="_vkyDAHbLEfGv-tGzRUrLbQ"
    sysmlQualifiedName=
    "QN::BB84_Network::QC_configuration::Distance"
    simulatorParameter="QC_Distance"
    value="1.0"
    unit="m"/>

<outputs
    xmi:id="_37J4EHbLEfGv-tGzRUrLbQ"
    simulatorMetric="mean_error_rate"
    resultFile="bb84_results.json"
    sysmlQualifiedName=
    "QN::BB84_Network::simulatedErrorRate::errorRate"/>

</qnctrl:SimulatorController>
\end{lstlisting}

The execution adapter reads the input bindings, assigns their values to the corresponding SeQUeNCe parameters, and executes the BB84 experiment shown in Listing~\ref{5th}. The experiment configures the channel distance and specifies a key size of 256 bits and a simulation end time. Although additional parameters and performance measures could be incorporated, this example varies only the quantum-channel distance to align with the physical QKD demonstration. The resulting \texttt{mean\_error\_rate} is exported to the JSON file, returned to the corresponding SysML v2 property identified by the output binding.

\begin{lstlisting}[
caption={SeQUeNCe BB84 channel configuration, execution, and result export.},
label={5th},
frame=single,
basicstyle=\ttfamily\footnotesize
]
qc = QuantumChannel(
    "qc0",
    tl,
    distance=QC_Distance
)
pair_bb84_protocols(
    alice.protocol_stack[0],
    bob.protocol_stack[0]
)
process = Process(
    alice.protocol_stack[0],
    "push",
    [256, math.inf, 6e12]
)
bb84 = alice.protocol_stack[0]

mean_error_rate = (
    sum(bb84.error_rates) / len(bb84.error_rates)
    if bb84.error_rates else None
)
results = {
    "protocol": "BB84",
    "distance_m": QC_Distance,
    "mean_error_rate": mean_error_rate
}
with open(
    "bb84_results.json",
    "w",
    encoding="utf-8"
) as file:
    json.dump(results, file, indent=4)
\end{lstlisting}

Although this feasibility example does not fully implement Architecture~(d), it demonstrates a viable model-driven path linking the SysML v2 hub, a physical QKD kit, and a corresponding QN simulation. The same network configuration is applied to both environments, enabling the test configuration to be assessed and the results to be returned to the hub. This aligns with the two loops in Figure~\ref{fig:Req}, which identify the DT requirements for design-time evaluation and runtime synchronisation. It also aligns with the capability of SysML v2 to support several tools, hardware, scenarios, and technologies. Future work is required to demonstrate the full implementation by automating the transformations, integrating additional simulators and domain models, and realising runtime synchronisation.

\section{Conclusions}

This paper examined the emerging use of digital twins for quantum networks and identified key requirements, including scalability, adaptability, interoperability, consistency, and traceability. It proposed progressive code-driven, domain-model-driven, and system-model-driven architectures, showing how MDE can make integration knowledge more explicit, reusable, and manageable. Future work will implement and evaluate the framework as an integrated QN digital twin with swappable simulator and hardware interfaces, enabling performance assessment across different protocols, physical effects, configurations, and operational scenarios.

\bibliographystyle{ACM-Reference-Format}
\bibliography{acmart}

\end{document}